\magnification \magstep 1
\hsize 6 true in
\vsize 8.5 true in
\input amssym.def
\input amssym.tex
\nopagenumbers
\openup 1 \jot
\centerline {\bf {TOPOLOGY CHANGE IN CLASSICAL AND QUANTUM GRAVITY}}
\vskip
1.5cm
\centerline {G W GIBBONS}
\centerline {D.A.M.T.P.}
\centerline {University of Cambridge}
\centerline {Silver Street}
\centerline {Cambridge CB3 9EW}
\centerline {U.K.}
\vskip 1.5cm
\centerline  {\bf {ABSTRACT}}
\tenrm
{\narrower In these two lectures I describe the difficulties
one encounters when trying to construct a framework in which to describe
topology change in classical general relativity where one sticks to the
assumption of an everywhere non-singular Lorentzian metric and how these
difficulties can be circumvented
in the Euclidean approach to quantum gravity. \smallskip}
\vskip .5 cm

\noindent

Originally circulated as Topology change in classical and quantum gravity.
G.W. Gibbons, (Cambridge U.) . Jun 1991. 27pp.
To appear in Proc. of 10th Sorak School of Theoretical Physics to be publ. 
by World Scientific, ed. by J.E. Kim and is an extended version of 
lecture given at 6th Marcel Grossmann Mtg., Kyoto, Japan, Jun 23-29, 1991.
Published in Mt. Sorak Symposium 1991:159-185 (QCD161:S939:1991) 
The two lectures appeared in print as 
Topology change in classical and quantum gravity 
{\it Recent
Developments in Field Theory} ed. Jihn E Kim ( Min Eum Sa, Seoul ) (1992)
The lecture in Kyoto appeared as  
Topology change in Lorentzian and Riemannian Gravity,
pp.  {\it Proceedings of the Sixth Marcel Grossmann Meeting: 
Kyoto 1991} eds. H Sato and T Nakamura (World Scientific, Singapore) 
1013-1032 (1992)

\vskip .5cm 

\noindent {\bf {Introduction}}

An important question both in classical and in quantum gravity theory is whether
the topology of space can change . In other words one may ask whether it is
possible for the 4-dimensional spacetime manifold $M$  not to have the product
topology
of the real line times some spatial 3-manifold but rather something more
complicated. Intuitively we are interested in processes
variously described as:

``trouser leg cosmology''

``The birth of the universe from nothing''

``The birth of twin universes''

``The creation of a wormhole in an $S^3$ universe''.

Of course ideas
this sort, involving as they do a definite three or four dimensional geometry
can  make sense {\bf {at best}} classically or at the semi-classical
level - no matter what the correct underlying theory of quantum gravity turns
out to be. Thus  by quantum gravity I shall mean semi-classical quantum gravity
and more specifically I shall be describing an instanton approach within the
framework of Euclidean Quantum Gravity.

Before turning in detail to the problem at hand it is worth pausing
to remind ourselves of the range
of validity of such semi-classical considerations.  We know that some sort of
new physics must set in at a scale which it is convenient to characterize by a
temperature.  Let us agree to call this temperature the Hagedorn temperature $T
_ {Hagedorn}$.  In string theory for example this is given approximately by
$(\alpha') ^ {- {1 \over 2}}$ when $\alpha'$ is the string tension.

Now ``conventional'' Einstein quantum gravity effects of the sort that have
been much discussed at this meeting using rough speaking semi-classical ideas
and expansions around classical solutions of possibly modified Einstein
equations set in at roughly the Planck temperature $T _ {Planck} = G ^ {- {1
\over 2}}$ where $G$ is Newton's constant.  In order to have a clear cut
separation between these effects and those due to new physics - for example
``stringy'' effects - we need that $T _ {Planck} / T _ {Hagedorn}$ should be
small, certainly less than unity and hopefully less than or of the order of $10
^ {-2}$, on the other hand, $T _{Planck}/ T _ {Hagedorn}$ is greater or of the
order of unity little of what I have to say in this talk, or indeed little of
much of the work in quantum gravity described at this meeting, makes much
physical sense.

In string theory as it is at present the relation between Newton's constant $G$
and the string tension $\alpha'$ is rather uncertain.  Optimistic attempts at
phenomenological models might work if $T _{Planck}/ T _ {Hagedorn}$ is roughly of
order $10$, but this is hardly conclusive.  The realistic position is that we
simply do not know $T _{Hagedorn}/T _ {Planck}$ - it might be greater than
unity, it might be less than unity. It is worth remembering that  the name $T _ {Hagedorn}$ stems from a
time when string theory was believed to be a theory of Hadron physics and its
value compared with $T _ {Planck}$ was thus tiny.  The main point to be borne
in mind is that all that I have to say here may or may not turn out to be
physically relevant when, or if, we finally discover what the new physics
really is.

With this caveat in mind here is the plan of what I want to cover:

\noindent {\bf {1. Topology Change in Classical Relativity}}. By this I mean
via a spacetime which carries an every-where non-singular Lorentz metric.
This is an old
topic. The basic conclusion is that this raises difficulties on purely kinematic
grounds because of the necessity of non-time-orientability or of closed
timelike curves. I shall also describe some new selection rules
which Stephen Hawking and I have recently discovered which show that in some
cases even if one is prepared to contemplate topology change via Lorentzian
metrics with closed timelike curves a potentially even more disastrous problem
arises, the impossibility of introducing 2-component spinors on the manifold
$M$.

\noindent {\bf {2.  Topology Change in Euclidean Quantum Gravity}}  By contrast
with an approach based on every-where non-singular Lorentzian metrics in which
one encounters great difficulties even at the purely kinematic level in
describing topology change there are no such difficulties in the Euclidean Path
Integral Approach. Not only is topology change kinematically allowed but
there are
actually classical complex paths in the path integral which mediate it. These
classical paths correspond to what Hartle and I have called Real Tunneling
Geometries. I will describe this
attempt to formalize
the idea of a complex classical path in the framework of the path integral
approach to Quantum Gravity give a number of examples
representing the
production of pairs of black holes by strong external electromagnetic or
cosmological fields. These classical solutions can  serve as the basis for an
Instanton calculation
of the amplitude. Quantum corrections can be treated in a variety of ways but
here I shall sketch an approach to
the
quantization of fluctuations about real tunneling geometries based on
Reflection Positivity which makes
contact with the ideas of Euclidean Quantum Field theory.

\vskip 1.5cm
\leftline {\bf {1.  Topology Change in Classical Gravity }}
By a classical topology change between to 3-manifolds $\Sigma_{initial}$ and
$\Sigma_{final}$  we mean  that there  exists a compact connected
4-dimensional spacetime $M$ whose boundary $\partial M= \Sigma_{ initial} \cup
\Sigma_{final} = \Sigma$ with an everywhere non-singular Lorentz metric with
respect to which the 3-manifolds are spacelike. Such a spacetime is called a
Lorentz-cobordism between the initial and final 3-manifolds
$\Sigma _{initial}$ and $\Sigma_{final}$ respectively.  Note that neither the
initial nor the final 3-manifold need be connected. In fact it is convenient
to consider the spacetime as having just one spacelike boundary $\Sigma$ which is the
union of all its components. We remark that the Lorentzian spacetime $M$ need not be, and
in general will not be, geodesically complete.
By Rohlin's theorem there are many compact 4-manifolds $M$ whose boundary
$\partial M = \Sigma$, where $\Sigma$ is a closed but not necessarily connected
oriented 3-manifold.  Topologically therefore topology change is always
possible.  However to have a spacetime we must endow $M$ with a Lorentzian
metric.

We can always give $M$ a Riemannian metric $g ^ R _ {\alpha \beta}$ with
signature $++++$.  By contrast however to give $M$ a Lorentzian metric $g ^ L _ {\alpha
\beta}$ with signature $+++-$ requires that the Euler characteristic $\chi(M)$
vanishes.  As an aside let me remark, particularly in the context of studies of
signature change, that the remaining possibility - signature $++--$ is worthy
of study and constitutes a sort of ``Last Frontier'' as far as 4-dimensional
geometry is concerned.  The analogous condition turns out to be that the Euler
characteristic $\chi$ should be even and equal to the Hirzebruch signature
$\tau$ modulo 4.

To return to the Lorentzian case.  We can diagonalize $g ^ L _ {\alpha \beta}$
with respect to some arbitrary purely axillary Riemannian metric $g ^ R _ {\alpha \beta}$ to
obtain a line field $\pm V,^\alpha$ where $V ^ \alpha$ is the eigenvector with
negative eigenvalue. Conversely given a line field  $\pm V^{\alpha}$ which we
way normalize with
respect to $g ^ R _ {\alpha \beta}$ we may construct a Lorentzian metric in
terms of
$$
V _ \alpha = g ^ R _ {\alpha \beta} V ^ \beta
\eqno (1)
$$
by:
$$
g ^ L _ {\alpha \beta} = g ^ R _ {\alpha \beta} - 2V _ \alpha \ V _ \beta.
\eqno (2)
$$

Time-orientability of the Lorentzian metric is equivalent to being able to
remove the $\pm 1$ ambiguity remaining in the definition of $V ^ \alpha$ and
obtaining a vector field $ {\bf V}$ on $M$.  To say that the boundary
$\partial M = \Sigma$ is spacelike with respect to $g ^ L _ {\alpha \beta}$
amounts to saying that the line field is transverse to $\Sigma$.  A theorem of
Hopf then implies that the necessary and sufficient condition for the existence
of the line field is the vanishing of the Euler characteristic
$\chi(M)$.  According to the results of Misner, Rhinehart, Geroch, Yodzis,
Sorkin etc this can always be achieved in 4-dimensions (but {\bf {not}} in
2-dimensions) by taking connected sums of the original manifold $M$, which may
not have vanishing Euler characteristic, with suitable closed 4-manifolds.
The connected sum $M_1 \# M_2 $ of two manifolds $M_1$ and $M_2$ is obtained by
removing a 4-ball from both and and gluing together by identify the two  $S^3 $
boundary components so created. Under connected sum we have:
$$\chi (M _ 1 \# M _ 2 ) = \chi (M _ 1) + \chi (M_2) - 2
\eqno (3)
$$

Thus for example:

\ \ \ \ \ \ $M \# S ^ 2 \times S ^ 2$ has $\chi$ increased by 2,

\ \ \ \ \ \ $M \# S ^ 1 \times S ^ 3$ has $\chi$ decreased by 2,

\ \ \ \ \ \ $M \#  {\Bbb C} {\Bbb P} ^ 2 $ has $\chi$ increased by 1,

\ \ \ \ \ \ $M \#  {\Bbb R} {\Bbb P}^4$ has $\chi$ decreased by 1.

We see that if $\chi$ is even we can use $S ^ 2 \times S ^ 2$ or $S ^ 1 \times
S ^ 3$ to reduce $\chi$ to zero.  These manifolds are familiar:  $S ^ 2 \times
S ^ 2$ corresponds to a (euclidean) black hole topology and $S ^ 1 \times S ^
3$ to a (euclidean) wormhole.  On the other hand if $\chi$ is odd we may have
to use less benign manifolds like $ {\Bbb C}{\Bbb P}^2$ or $ {\Bbb R}{\Bbb P}^4$ to reduce $\chi$ to
zero.  The former has no spin structure and the latter is not orientable.  We
shall see that this can give rise to problems.

As an {\bf {example}} consider the case of $\Sigma = S ^ 1 \times S ^ 2$ the
3-dimensional wormhole beloved of Wheeler.  Since
$$
\Sigma = S ^ 1 \times S ^ 2 = \partial (S ^ 1 \times B ^ 3)
\eqno (4)
$$
where $S ^ 1$ has angular coordinate $\psi$ such that $0 \leq \psi \leq 2 \pi$
and the 3-ball $B^3$ has polar coordinates ($t, \ \theta, \ \phi$) with $0 \leq
t \leq 1$, $0 < \theta \leq \pi$, $0 < \phi \leq 2\pi$.  We can give $S ^ 1
\times B ^ 3$ the flat Riemannian metric
$$
dS^2_R = d \psi ^ 2 + dt ^ 2 + t ^ 2 (d \theta ^ 2 + sin ^ 2 \theta d \phi^2)
\eqno (5)
$$
and suitable vector field:
$$
{\bf V} = b(t) {\partial \over \partial \psi} + a(t) {\partial \over
\partial t}
\eqno (6)
$$
where $a ^ 2 + b ^ 2 = 1$ and $a$ passes smoothly and monotonically from 0 to 1
as $t$ runs from 1 to 1, with vanishing derivative at $t = 0$.  The resulting
Lorentzian metric:
$$
ds^2_L = d\psi^2 + dt ^ 2 + t^ 2 (d \theta ^ 2 + \sin ^ 2 \theta d \phi ^ 2 ) -
2 (b d \psi + a \ dt ) ^ 2
\eqno (7)
$$
is non-singular but in general geodesically incomplete.  If $t _ c$ is such
that $a (t_c) = {1 \over \sqrt {2}}$ then $t = t _c$ is a Cauchy horizon and
for $0 < t < t _ c$ the spacetime contains closed timelike curves.
Nevertheless we can claim from this example that {\it {the creation of a worm
hole from nothing via a time orientable spacetime is certainly an allowed
process at the purely kinematic level.}}

I did not specify earlier which components of $\partial M$ lay in the past and
which in the future, if $M$ {\bf {is}} time orientable.  This is because of my
next example which shows how to turn round the direction of time.  Let $M =
\Sigma \times I$ where the 3-manifold $\Sigma$ has metric $g _ {ij} (x ^ k) , \
k = 1, 2, 3$ and the interval $I$ has coordinate $t, \ -1 \leq t \leq +1$.
Choose as Riemannian metric on $M$ the product metric
$$
ds ^ 2 _ R = g _ {ij} (x) dx ^ i dx ^ j + dt^2
\eqno (8)
$$
and if ${\bf U} (x ^ i)$ is a unit vector field on $\Sigma$ (which always
exists in 3-dimensions),
choose for the vector field ${\bf V}$:
$$
{\bf V} = b(t) {\bf U} + a (t) {\partial \over \partial t}
\eqno (9)
$$
where $a(t)$ is the same function as used earlier but extended to $-1 \leq t
\leq 0$ by the requirement that $a(t) = -a (-t)$, i.e. it be an odd function of
$t$.  Thus ${\bf V} $ is outgoing on both components of the boundary of
$\partial M = (\Sigma \times 1) \cup (\Sigma \times -1)$.  Again $M$ has Cauchy
horizons and closed timelike curves, and is incomplete in general.
Nevertheless it shows that we need not specify the direction of time on any
component of the boundary of a Lorentz cobordism because by attaching a
suitable copy of this product manifold to that component we can change the
future to the past or vice-versa.

We can also claim that {\it {pure kinematically the birth of twin universes
from nothing is possible in the Lorentzian picture}}.

The traditional problem with these Lorentz-cobordisms, which will in general be
geodesically incomplete, is encapsulated in the well known theorem of Geroch [1] to the effect
that either

(1) there is no global time orientation or

(2) they contain closed timelike curves.

We have seen examples of (2), to see an example of (1) consider $\Sigma = S ^
3$ with $M =  {\Bbb R}{\Bbb P} 4 - \{pt\} \sim S ^ 3 \times {\Bbb R} $.
The metric
could be of time-symmetric $F-L-R-W$ form:
$$
ds _ L ^ 2 = - dt ^ 2 + R ^ 2 (t) (d \psi ^ 2 + \sin ^ 2 \psi (d \theta ^ 2 +
\sin ^ 2 \theta \theta \phi ^ 2) )
\eqno (10)
$$
with $(t, \ \psi, \ \theta, \ \phi)$ identified with $(-t, \ \pi - \psi, \ \pi -
\theta , \ \phi + \pi)$ and where $R (t) = R (-t)$.
If we restrict $t$ to $(-1 \leq t \leq + 1)$ we have just one boundary
by virtue of the
identification.  Since the identification involution, call it $J$, reverses the
sense of time the quotient is not time-orientable.  However the quotient does
{\bf {not}} contain any closed timelike curves since the 2-fold covering
manifold does not contain any.  Note that by a closed timelike curve we mean a
closed curve whose tangent vector always points in the {\sl same half} of the
light
cone. Thus if the worldline suffers  a jump in velocity at some point the two
tangent vectors at that point must lie in the same half of the light cone at
that point. In fact it can happen in our example that two points identified
under the involution $J$ are timelike separated on the 2-fold covering
manifold. However because of the reversal of time orientation these give a
closed curve which sets of from some point into the future and returns from the
future, i.e in the same half of the light-cone. If $R(t) = (3/\Lambda) ^ {1 \over 2} \cosh
(({\Lambda \over 3}) ^ {1 \over 2} t ) $ we obtain deSitter Spacetime and we
are talking about the so-called ``elliptic interpretation''.  The involution
$J$ is then just the antipodal map. In De-Sitter spacetime a point and its antipode
are never timelike separated, so we don't have pathological closed curves of the
sort described above in this case.  Thus we can claim {\it {that the elliptic
interpretation corresponds to the birth from nothing of a single $S^3$
universe.}}

The lack of time-orientability leads to difficulties with introducing spinors
on the Lorentzian spacetime and also with quantizing fields on the spacetime.
One way of expressing this latter difficulty is to say that one is lead to
consider real quantum mechanics.  More geometrically the time-reversing
involution $J$ induces an anti-symplectic involution on the phase space of
classical fields, i.e. on the space of Cauchy data.  Thus attempts to quantize
using conventional ideas are stymied.  These difficulties are detailed in [2].

The question then arises can one find a time-orientable Lorentz-cobordism for a
single $S^3$ universe.  This is possible, by removing a point from
$ {\Bbb C}{\Bbb P}^ 2
\# S ^ 1 \times S ^ 3$ for example.  However this cobordism is not a spin
manifold.  In fact this is a general problem:  {\it {No Lorentz-cobordism for
$S^3$ can admit an $SL(2,  {\Bbb C})$ spinor structure.}}

The {\bf {proof}} is simple.  If $M$ admits an $SL(2,  {\Bbb C})$ spinor structure it
must also admit a Spin (4) structure - i.e. be a spin manifold in the
conventional sense, with vanishing second Stiefel-Whitney class $w _ 2 \in H ^
2 (M; \ {\Bbb Z})$.  We can fill in the boundary $\partial M = S ^ 3$ with a 4-ball
$B^4$ to get a closed 4-manifold $\tilde M$.  The spin(4) structure extends to
$\tilde M$, which moreover has Euler characteristic $\chi (\tilde {M}) = 1$.
We have thus constructed a closed spin 4-manifold with odd Euler
characteristic.  However:

{\bf {Lemma}}:  Every closed spin 4-manifold has even Euler characteristic.

Thus we obtain a contradiction.  To prove the lemma note that:
$$
\chi (\tilde {M}) = 2 - 2b_1 + b_2^+ + b_2 ^ -
\eqno (11)
$$
where $b_1$ is the first Betti number of $\tilde {M}$ and $b _ 2 ^ +, \ b _ 2 ^
-$ the dimension of the spaces of harmonic 2 forms which are self-dual,
respectively antiself-dual.  Using results of Hirzebruch, Atiyah and Singer one
has that the index of the Dirac operator $D$ on $\tilde {M}$ with respect to
some (and hence every) Riemannian metric on $\tilde {M}$ is given by
$$
\rm {index} \ {D} = (b _ 2 ^ + - b _ 2 ^ - / / 8
\eqno (12)
$$

On a closed 4-manifold index ($D$) is even and thus from (2.11) we see that
$\chi$ must be even.

Extending the argument in an obvious way we obtain a new {\it {selection rule for
$S^3$ universes.}}

The number of $S^3$ universes in any Lorentz cobordism admitting an $SL(2,  {\Bbb C})$
spinor structure is conserved modulo 2. Another way to express this is to say
that there is a
${\Bbb Z}_2$ {\bf {topological invariant}} $u(\Sigma)$ for closed
orientable  3-manifolds such that
$$
u(\Sigma) = 0
\eqno (13)
$$
if $\Sigma$ admits a spin-Lorentz-cobordism and
$$
u (\Sigma) = 1
\eqno (14)
$$
otherwise.

In fact one may prove that $u(\Sigma)$ behaves exactly as one expects of a
conserved quantum number under
disjoint union
$$
u (\Sigma _ 1 \cup \Sigma _ 2) = u (\Sigma _ 1) + u (\Sigma _ 2) \ \rm {mod} \ 2
\eqno (15)
$$
while under connected sum it satisfies
$$
u (\Sigma _ 1 \# \Sigma _ 2) = u (\Sigma _ 1) + u (\Sigma _ 2) + 1 \ \rm
{mod} \ 2
\eqno (16)
$$

From the discussion above $u (S^3) = 1$ and our previous example for
$S ^ 1 \times S ^ 3$ (which
admits spin for both choices of spin structure on the boundary) shows that $u
(S ^ 1 \times S ^ 2) = 0$.  This shows that {\it {single wormholes cannot be
created in the laboratory - they must be created in pairs.}}  In other words if
we start with no wormholes, i.e. $\Sigma _ {initial} = S ^ 3$ and end with the
connected sum of $k \ S ^ 1 \times S ^ 2$'s, $\Sigma _ {final} = \# _ k \ S
^ 1 \times S ^ 2$, then since using these rules:
$$
u (\Sigma _ {initial} \cup \Sigma _ {final} ) = 1 + (k - 1) \ \rm {mod} \ 2
$$
we must have $k = 0 \ \rm {mod} \ 2$ if our ``laboratory'' is to allow $SL(2, \
 {\Bbb C})$ spinors.

One may express $u(\Sigma)$ in terms of the ${\Bbb Z}_2 $
homology groups of the boundary. It turns out to be the mod 2 Kervaire
semi-characteristic, i.e.
$$
u (\Sigma) = \dim _ {{\Bbb Z}_2} (H  _ 0 (\Sigma ; \ {\Bbb Z}_ 2 )
\oplus H_1
(\Sigma ; {\Bbb Z} _ 2)) \ \rm {mod} \ 2
$$
where $H_i ( \Sigma ; {\Bbb Z} _ 2)$ is the i'th cohomology group of $\Sigma$
with ${\Bbb Z}_2$ coefficients.  Thus for
$ {\Bbb R}{\Bbb P} ^ 3 \cong SO(3)$ we have $u( {\Bbb R}{\Bbb P}^3) =
0$ and this selection rule does not prevent such universes being born from
nothing.  More details will be given in a forthcoming paper with Stephen
Hawking.

Before concluding this section I think it might be worth mentioning some
results of Thurston
on foliations which may be applied to the present case. Firstly the vanishing
of the Euler characteristic of a closed manifold is  the neccessary and
sufficient condition for the existence of a smooth foliation by co-dimension one
leaves. If the manifold $M$ has a boundary, which is more direct interest to us
then Thurston proves that  it admits a smooth co-dimension one foliation
tangent to the boundary $\partial M$ if and only if it admits a line field
transverse to the boundary and either

1) $M = \Sigma \times I$    or $M= \Sigma \times _{{\Bbb Z}_2} I$ where ${\Bbb
Z}_2$ acts as the involution $J$ in our previous examples

\noindent or

2) For each boundary component $\partial M _i$ ,
$ H^i (\partial M_i ; {\Bbb R})$
is non-trivial.

\noindent Condition 2) may be paraphrased as saying that that there must be a wormhole
of some sort on each boundary
component. Thus, according to Thurston's theorem the cobordism for $S^1 \times
S^2 $ used above admits such a foliation.
Superficially one might suppose that one could use these foliations to
construct a Hamiltonian  description of topology change in the Lorentzian
regime but  it must be borne in mind that the leaves need not all be
diifeomorphic, and even if they are there will not exist a global time
function.  Nevertheless it would be interesting to investigate such foliations
in more detail.
\vskip 1.5cm
\noindent {\bf {2. Topology Change in Quantum Gravity}}

The results described above for spin-Lorentz-cobordisms should be contrasted
with the by now familiar picture of the creation of the universe from nothing
obtained using ideas from quantum tunneling.  Hartle and I [3] have formulated a
general framework into which almost all known examples seem to fit,
with the
exception that our compactness assumption may need to be relaxed.  The idea is
to seek solutions of the relevant classical field equations (i.e. saddle points
in the functional integral) associated with a (closed) 3-manifold $\Sigma$ with
3-metric $h _ {ij}$ and

(1) a (compact) Riemannian 4-manifold $M _ R$ with boundary $\partial M _ R =
\Sigma$.

(2) a Lorentzian 4-manifold $M _ L$ for which $\Sigma$ is a (partial) Cauchy
surface

\noindent such that

(3) $\Sigma$ is totally geodesic with respect to both $M _ R$ and $M _ L$, that
is its second fundamental form $K_{ij} = 0$.

\noindent The motivation for this idea comes from the instanton or "bounce"
description
of

1) False Vacuum Decay

\noindent or

2) Pair-creation by strong electric fields (Schwinger Process)

\noindent In both cases the stationary point of the classical Euclidean action has a
symmetry under the reversal of imaginary time. The behaviour after tunneling is
described by a solution of the classical equations in real time whose initial
values coincide with the values of the solution in imaginary time at imaginary
time zero. By joining the imaginary time trajectory for negative values of
imaginary time to the real time trajectory for positive values of real time one
obtains a complex classical path whose action will be complex, the imaginary
part coming from the imaginary time portion, the real part coming from the real
time portion. There is no contribution from the mid-point at time zero because
the velocities there vanish.

In General Relativity initial values correspond to giving the
3-metric on a spacelike surface $\Sigma$ and its second fundamental form
$K_{ij}$ (i.e.
its time derivative). If the data admit a moment of time symmetry at $\Sigma$
the second fundamental form must vanish. In this case the boundary data may
serve either as Cauchy data for the hyperbolic Lorentzian Einstein equations
or as Dirichlet data for the elliptic Riemannian Einstein equations.

The basic gravitational example is for $\Sigma = S ^ 3$.  $M _ R$ is half of $S ^ 4$ ie.
$$
ds ^ 2 _ R = d \tau ^ 2 + ({3 \over \Lambda}) \cos ^ 2 (\tau \sqrt{{\Lambda \over
3}})  \ d \Omega _ 3 ^ 2
\eqno (17)
$$
with $- {\pi \over 2} \leq \tau \sqrt {{\Lambda \over 3}} \leq 0$, and $M _ L$
is half of deSitter spacetime, i.e.
$$
d s _ L ^ 2 = - dt ^ 2 + ({3 \over \Lambda}) \cosh ^ 2 (t \sqrt {{\Lambda \over
3}} ) \  d \Omega ^ 2 _ 3
\eqno (18)
$$
with $t \geq 0$.
Note that while the 4-metric is degenerate on the signature
changing surface $\Sigma$, given in our example by $\tau = 0 = t$, considered
as a curve of 3-metrics the vanishing of the second fundamental form on
$\Sigma$ means that there are no contributions to the variation of the action
functional at $\Sigma$ because the curve suffers no jump in slope.  Put
another way there is no distributional contribution to the Einstein tensor at
$\Sigma$.  Thus real tunneling geometries may be considered as true critical
points of the classical action functional.  The compactness of $M _ R$ is
convenient, for example when studying the No Boundary proposal of Hartle and
Hawking [4] but not, I feel, an essential part of the idea.  There are interesting
examples involving black holes for which $\Sigma$ is not closed and $M _ R$ is
not compact as we shall see.  As an aside:  let me remark that as models of
signature change rather than of tunneling one might also consider replacing $M
_ R$ by a manifold with an ultra-hyperbolic metric of signature $++--$, in
which case $\Sigma$ would have a timelike metric.  There is, I believe, an
argument for calling metrics with as many positive directions as negative
directions ``Kleinian''.  Much of the present theory goes through if $R$ is
replaced by $K$ under the understanding that $K$ stands for Kleinian in that
sense.  Because Kleinian 4-metrics admit the idea of self-duality and because
they arise in some (rather exotic) string theory this ``last frontier'' seems
ripe for colonization.  Some pioneering attempts will be contained in [5].
Let me return however to the case of Riemannian tunneling.

Associated with the Riemannian manifold $M _ R$ and the Lorentzian manifold $M
_ L$ are their doubles $2M _ R$ and $2M _ L$ respectively which are obtained by
joining two copies of $M _ R$ on $M _ L$ respectively across their common
boundary $\Sigma$.  Thus $2M_L$ is a spacetime admitting a ``moment of time
symmetry'' while $2M_R$ is a compact Riemannian manifold admitting a isometric
involution $\theta$ (or reflection map) which interchanges the 2 halves and
fixes $\Sigma$, i.e.
$$
2M _ R = M ^ + _{R} \cup  M _ R ^ -
\eqno (18)
$$
with
$$
\theta M ^ {\pm} _ R = M ^ {\mp} _ R ,
\eqno (19)
$$
and thus
$$
\theta \Sigma = \Sigma
\eqno (20)
$$
$$
\theta ^ 2 = id.
\eqno (21)
$$

In fact $2M_R$ and $2M_L$ may be regarded as two real slices of the
complexified manifold $M _
c$
of complex dimension 4 which carries a symmetric complex covariant tensor
field of type $(2, \ 0)$ which restricts on $2M _ R$ and $2M_L$ to the real
metrics $g ^ R _ {\alpha \beta}$ and $g ^ L _ {\alpha \beta}$ .  The real slices
$2M_R$ and $2M_L$ are stabilized by two anti-holomorphic involutions acting on
$M _ c$, $J _ R$  and $J _ L$ respectively.  Thus
$$
J _ R (2M _ R) = 2 M _ R
\eqno (22)
$$
$$
J _ L (2M _ L) = 2 M _ L
\eqno (23)
$$
Restricted to $M _ R \ \ J _ L$ coincides with our previous reflection map
$\Sigma$ and $J _ R$, restricted to $M _ L$ corresponds to time-reversal.

As an example one may consider de Sitter spacetime as a complex quadric in
$ {\Bbb C}
^ 5$.  This is described in the article with Hartle.  Here I will give a
different example:  the Schwarzschild solution.  This may be complexified as an
algebraic variety in $ {\Bbb C} ^ 7$ with complex coordinates
$Z ^ {\alpha}, \ \alpha = 1,
\dots 7$ [6].  In terms of local Schwarzschild coordinates (which cover only a
portion of the variety)
$$
Z ^ 1 = r \sin \theta \cos \phi
\eqno (24)
$$
$$
Z ^ 2 = r \sin \theta \sin \phi
\eqno (25)
$$
$$
Z ^ 3 = r \cos \theta
\eqno (26)
$$
$$
Z^ 4 = -2M ({2 M \over r}) ^ {1 \over 2} + 4M ({r \over 2M}) ^ {1 \over 2}
\eqno (27)
$$
$$
Z ^ 5 = 2M \sqrt {3} ({2M \over r}) ^ {1 \over 2}
\eqno (28)
$$
$$
Z ^ 6 = 4M (1 - {2M \over r}) ^ {1 \over 2} \cosh (t /4M)
\eqno (29)
$$
$$
Z ^ 7 = 4M (1 - {2M \over r}) ^ {1 \over 2} \sinh (t/4M)
\eqno (30)
$$
The algebraic variety is given by the 3-equations
$$
(Z ^ 6) ^ 2 - (Z ^ 7) ^ 2 + {4 \over 3} (Z ^ 5) ^ 2 = 16M ^ 2
\eqno (31)
$$
$$
((Z ^ 1) ^ 2 + (Z ^ 2)^2 + (Z ^ 3) ^ 2) (Z ^ 5) ^ 4 = 576M ^ 6
\eqno (32)
$$
$$
\sqrt {3} Z ^ 4 Z ^ 5 + (Z ^ 5) ^ 2 = 24 M ^ 2
\eqno (31)
$$

The Lorentzian section $2M _ L$ is stabilized by
$$
J _ L: (Z ^ 1, Z ^ 2, Z ^ 3, Z ^ 4, Z ^ 5, Z ^ 6, Z ^ 7) \rightarrow (\overline
{Z} ^ 1, \overline {Z} ^ 2, \overline {Z} ^ 3, \overline {Z} ^ 4, \overline {Z}
^ 5, \overline {Z} ^ 6, \overline {Z} ^ 7)
\eqno (32)
$$
and the Riemannian section $2M _ L$ by
$$
J _ R:(Z ^ 1, Z ^ 2, Z ^ 3, Z ^ 4, Z ^ 5, Z ^ 6, Z ^ 7) \rightarrow (\overline
{Z} ^ 1, \overline {Z} ^ 2, \overline {Z} ^ 3, \overline {Z} ^ 4, \overline {Z}
^ 5, \overline {Z} ^ 6, - \overline {Z} ^ 7)
\eqno (33)
$$
These intersect on the familiar 2-sheeted ``Einstein Rosen bridge'' $\Sigma$
with topology $S ^ 2 \times {\Bbb R} $ given by $Z ^ 7 = 0$.  In terms
of the real time coordinate $t, \ \Sigma$ is given by $t = 0$.  In terms of the
imaginary time coordinate $\tau = it$, which is periodic with period $8 \pi M$,
$\Sigma$ is given by $\tau = 0$ and $\tau = 4\pi M$.

In a tunneling context the Schwarzschild solution has been applied to the
instability of hot flat space [7].  Of course neither $M _ R$ nor $\Sigma$
is compact in the present case but physically this is not unreasonable since in
practice one would be considering only a large but finite volume of spacetime
and intending to compute a tunneling probability per unit time per unit volume,
so at some stage in the calculation one would have to take a suitable ratio to
get a finite answer but this is a technicality which I won't dwell on here.

Having isolated the essential features of the geometries of interest for
tunneling one can proceed to investigate their properties in a systematic way.
Some results in this direction are given in the paper with Hartle.  For the
moment I will restrict attention to just two results.  One is what we called
the ``{\bf {Unique Conception Theorem}}''.  Suppose $M _ R$ is compact and the
Ricci tensor of the Riemannian metric $g ^ R _ {\alpha \beta}$, is, considered
as a quadratic form on tangent vectors, non-negative.  Then the boundary
$\Sigma$ must be connected.  That is one cannot find classical solutions
representing the birth from nothing of more than one universe if the positivity
restriction on the Ricci curvature holds.  It should be pointed out that there
are perfectly reasonable Lagrangians for which the Ricci tensor does not
satisfy these restrictions.  In fact the mathematical fact behind the theorem
is the same fact that allows one to prove that there are no 4-dimensional
Riemannian solutions representing ``wormholes'' subject to an appropriate
restriction on the Ricci curvature [8].  As is well known by considering
axions etc such wormhole solutions can be found and similarly one could find
solutions which nucleate the birth of more than one universe by considering
axion fields.

The second result is on the topology of connected components of $\Sigma$.
Again under suitable Positivity assumptions, the most important being a
positive rather than negative cosmological constant, it follows that the Ricci
scalar of $\Sigma$ must be positive.  From this one may, following the work of
Schoen, Yau and others deduce restrictions on the topology of $\Sigma$.  The
details are given in the paper with Hartle.  Much of the motivation of that
paper derived from quantum cosmology.  However the formalism also covers
the case of more localized topological fluctuations, as well as the ideas behind
false vacuum decay.  In the next section I shall describe some more examples
relating topology change to black hole theory.  Before doing so I want to
mention a new result which is not contained in the paper with Hartle
concerning the topology of homogeneous real tunneling metrics, i.e. those for
which the double $2M _ R$ admits the transitive isometric action of some Lie
group.

It seems plausible that the solutions of any set of field actions with least
action are homogeneous.  By Bishop's theorem [9] this is certainly true for
Einstein metrics with $\Lambda > 0$.  According to the classification by
Ishihara
[10] of 4-dimensional homogeneous Riemannian manifolds the universal
covering space $2
\tilde {M_R}$ of $2M _ R$ must be homeomorphic to one of ${\Bbb R}^4,
S ^ 4,  {\Bbb C} {\Bbb P}^2, S ^ 2 \times S ^ 2 ,{\Bbb R} \times S ^ 3$ and
$S
^ 2
\times {\Bbb R}^2 $.  In fact the existence of the reflection map
$\theta$ rules out $ {\Bbb C}{\Bbb P}^2$ (since it admits no orientation reversing
diffeomorphism).  {\it {Thus, quite independently of any field equations, the
topological possibilities for homogeneous real tunneling geometries are rather
limited.}}

As I mentioned above, for the Einstein equations with $\Lambda > 0$, the lowest
action solution is $S ^ 4$ with its standard homogeneous metric.

I shall now give three striking examples of
real tunneling geometries which support the idea that topology change,
involving black holes, does occur at the semi-classical level.  The three
examples are derived from:

(1) The Nariai - $S ^ 2 \times S ^ 2$, instanton

(2) The Mellor-Moss instanton

(3)  The Melvin-Ernst instanton.

The word ``instanton'' is synonymous with ``complete non-singular Riemannian
solutions of the classical field equations'', and refers in the present case to
the double $2M _ R$.

(1) {\bf {The Nariai Instanton}} is just the standard Einstein metric in $S ^
2 \times S ^ 2$ with its product metric.  The reflection map $\theta$ is just
reflection in a meridian of the first $S ^ 2$ factor.  Thus $\Sigma \equiv S ^
1 \times S ^ 2$, the $S ^ 1$ factor being made up of a pair of meridians.  The
Nariai metric may thus be thought of as nucleating the birth of an $S ^ 1
\times S ^ 2$ universe.  In view, however, of the horizons present in the
Lorentzian section, i.e. 2-dimensional De Sitter spacetime $\times S ^ 2$ I
would also like to view it as the creation of a pair of black holes from a
background cosmological field.  Something like this interpretation has been
given already by Perry and Perry and Ginsparg [11].  The interpretation
gains support from comparison with the Mellor-Moss case to be described later.
The basic idea is that a positive cosmological term causes pairs of particles
to separate because of the mutual repulsion they experience.  This repulsion
becomes larger the further the particles are apart.  Thus one expects a
positive cosmological term to give rise to a pair creation and of course for
conventional point particles this effect has been well understood for some
time.  Given that understanding it is reasonable to extend the idea to black
holes.  Now the Schwarzschild-De Sitter spacetime can be interpreted as
containing two black holes in a background De Sitter universe.  To do so one
must identify points in the Penrose diagram so that each surface of constant
time has one minimal 2-sphere and one maximal 2-sphere.  The cosmological
horizons intersect on the maximal 2-sphere and the black hole horizons on the
minimal 2-sphere.  The resulting spacetime has two static regions, each bounded
by a cosmological horizon and a black hole horizon.  One may regard the Nariai
metric as a limiting case of the Schwarzschild-De Sitter spacetime as $9 M^2
\Lambda  \ge 0$.  The black hole horizons are as large as they can be and
the cosmological horizons as small as they can be.  In this limit the metric
becomes a product metric and the two types of horizon becomes equivalent.

(2) {\bf {The Mellor-Moss Instanton}}  This is a particular member of the
Reissner-Nordstrom de Sitter family of solutions of the Einstein-Maxwell
equations with positive cosmological constant in which the mass parameter $M$
and the charge parameter $g$ (assumed purely magnetic for simplicity) are
related by [12]
$$
M = g/ \sqrt {4 \pi G}
\eqno (34)
$$
(electromagnetic units are ``rationalized'').  The surface gravities of the
cosmological and black hole horizons coincide in this limit and become:
$$
\kappa = \sqrt {\Lambda \over 3} (1 - 4M \sqrt {\Lambda \over 3}) ^ {1 \over 2}
\eqno (35)
$$

Thus the temperature of the two horizons is not zero, as it would be if $\Lambda
= 0$, but it is less than the temperature of De Sitter spacetime.  The
Riemannian metric has an imaginary time coordinate $\tau = it$ which is
periodic with period ${2\pi \over \kappa}$ and thus $2M_R$ is topologically $S ^ 2
\times S ^ 2$ but with a warped product metric.  The hypersurface $\Sigma$ has
topology $S ^ 1 \times S ^ 2$ as before and corresponds to the meridians $\tau
= 0$ and $\tau = \pi / \kappa $.  On analytic continuation to $M _ L$ these two
meridians are associated with two static regions each bounded by a black hole
horizon and a cosmological horizon.  A common feature of the Mellor-Moss, the $
S ^ 2 \times S ^ 2$ and the Schwarzschild cases is the necessity to consider
$\Sigma$ as containing two ``meridians'', $\tau = 0$ and $\tau = \pi /\kappa$.

The difference in action between the Mellor-Moss instanton and that of De
Sitter spacetime (for the doubles $2M_R$ in both cases) is
$$
M / T _ {De Sitter}
\eqno (36)
$$
where $T _ {De Sitter} = {1 \over 2 \pi} \sqrt {{\Lambda \over 3}}$.  Thus the
probability of tunneling is proportional to:
$$
\rm {Probability} \ \propto \ \exp - M/T _ {De Sitter}
\eqno (37)
$$
This seems an eminently reasonable answer.  Note that the probability is
proportional to $\exp - Action (2M_R)$ because it is the (modulus)$^2$ of the
amplitude which is proportional to $\exp - Action (M _R)$.

The special properties of extreme Reissner-Nordstrom black holes are widely
recognized and have been much studied.  In many ways they behave like solitons
in more familiar flat space field theories.  It seems very natural that they
should be created in pairs by strong external fields, including cosmological
fields.  Heuristic phase space arguments indicate that the most likely process
is that which entails the creation of particles with the least energy
consistent with the conservation of any relevant charges.  In the present case
the relevant charge is electromagnetic.  It is known that the extreme holes
satisfy a Bogomolnyi bound [13] and thus a condition like (4.1) is no
surprise, though, strictly speaking, it does not imply that the created black
holes have zero temperature.  My next example adds further weight to this
interpretation.

3. {\bf {The Melvin-Ernst Instanton}}  This is a solution of the
Einstein-Maxwell equations representing a pair of charged black holes in a
background Melvin type electromagnetic field.  For convenience, as with the
Mellor-Moss case we shall restrict attention to the purely magnetic case.  The
electrically charged case will also go through with minor modifications.  In the
absence of the background Melvin type field the solution is a special case of
the charged $C-$metrics discussed many years ago by Kinnersley and Walker [
14].
It is known that these metrics are singular - they contain ``nodal''
singularities.  This is to be expected.  If they did not one would be able to
construct real tunneling geometries and hence probability amplitudes for the
decay of flat Minkowski spacetime to a pair of black holes.  This contradicts
all our intuition, largely based on the Positive Mass theorem concerning the
stability of Minkowski spacetime (a proof of the positive mass theorem in the
presence of charged black holes may be found in [13]).  Examination of the
$C-$metrics does indeed reveal that they have vanishing ADM mass.  The
$C-$metrics contain both a black hole event horizon and an acceleration or
Rindler horizon.  In general their surface gravities and hence
their temperatures differ.  Thus even if they were free of nodal singularities
analytic continuation to give a non-singular Riemannian metric which is
periodic in imaginary time with a single period is problematic.

In more detail the charged C-metric has the form:
$$
ds ^ 2 = {1 \over A ^ 2 (x +y ) ^ 2} [ {d y ^ 2 \over F (y)}
 + {dx ^ 2 \over G (x)} + G (x) d \alpha ^ 2    -F(y)dt^2 ]
\eqno (38)
$$
where
$$
G (x) = - F (-x)
\eqno (39)
$$

$$
= 1 - x ^ 2 - 2GMAx^3-G (g^2/4 \pi ) A^2 x^4
\eqno (40)
$$
$$
=- G(g^2/ 4 \pi ) A^2 ( x - x_1)(x - x_2)(x- x_3) (x- x_4)
\eqno (41)
$$

I have labelled the 4 real roots of $G (x), \ x_1 \ x _ 2, \ x _ 3,  \  x _ 4$
in ascending magnitude ($ x_1$, $ x _ 2$ and $x _ 3$ are  negative and $x _ 4$ is
positive).

The range of the ``radial'' variable $y$ is
$$
-x _ 3 \leq y \leq - x _ 2
\eqno (42)
$$

with $y = |x _ 3|$ being an acceleration horizon and $y = |x _ 2|$ a black
hole horizon.  The range of the ``angular'' variable $x$ is $x _ 3 \leq
\alpha \leq x _ 4$.  The 2-surfaces $x = x _ 3$ and $x = x _ 4$ are axes of
symmetry for the angular Killing vector ${\partial \over \partial \alpha}$ .

In order to understand what the coordinates used it is helpful to consider the
case when the the mass parameter $m$ vanishes. Then the metric is flat and one
may transform to flat
inertial coordinates using the formulae:
$$
X^1  \pm iX^2 = {{(1-x^2)^{1 \over 2}} \over {A(x+y)}} \exp (\pm i \alpha)
\eqno (43)
$$

$$
X^3 \pm X^0 = {{(y^2-1)^{1 \over 2}} \over {A(x+y)}} \exp (\pm t )
\eqno (44)
$$

Evidently the coordinate singularity at $x=\pm 1$ is a rotation axis while
the coordinate singularity at $y= \pm 1$ corresponds to a pair of intersecting
null hyperplanes forming the past and future event horizons for a family of
uniformly accelerating worldlines. The points for which $x+y=0$ correspond to
infinity. A similar interpretation may be given in the
case  that $M \neq 0$ but there is in addition a Black Hole horizon. A detailed
description was given by Kinnersley and Walker [14].

If $0 \leq \alpha \leq \Delta \alpha$ there will be angular deficits:
$$
{\delta _4 \over 2 \pi} = {\Delta \alpha - \Delta \alpha _4 \over \Delta
\alpha _4} \ \ \ ; \ \ \ \ {\delta _ 3 \over 2 \pi} = {\Delta \alpha -
\Delta \alpha _ 3 \over \Delta \alpha _ 3}
\eqno (45)
$$
where
$$
\Delta \alpha _ 4 = {4 \pi \over |G ^ {\prime} (x_4 )|}; \thinspace \thinspace \thinspace  \Delta
\alpha _ 3 = {4 \pi \over |G ^ {\prime} (x_3)|}
\eqno (46)
$$
Since (unless $MA = 0$) $\Delta \alpha _ 4 \not = \Delta \alpha _ 3$ it is not
possible to eliminate both of these by choosing $\Delta \alpha$.  One can
eliminate $\delta _ 3$ in which case the black hole is pulled along by a
string, or $\delta _ 4$ in which case it is pushed along by a rod.

However it is a striking fact [15,16,17] that if condition (4.1) holds then
the two
surface gravities become equal.  This is not in itself a sufficient condition
to provide a regular instanton because the problem of the nodal singularity
remains.  However some years ago Ernst [18] showed,
using exact solution generating
techniques, how this nodal singularity could be eliminated by appending a
suitable electromagnetic field whose value is determined physically by the
condition that the acceleration induced by the electromagnetic field equals the
force needed to acceleration a massive black hole.  If no black hole is present
the relevant solution is ``Melvin's magnetic universe''.  If a black hole is
present the solution is asymptotic to Melvin's solution, and the strength of the
applied field is unconstrained. If an accelerating charged black hole is present
the strength of the appended field is determined in terms of the mass, charge
and acceleration parameters of the solution.

The Melvin solution represents an infinitely long
straight self-gravitating Faraday flux tube in equilibrium, the gravitational
attraction being in equipoise with the transverse magnetic pressure.
The metric is:
$$
ds^2 = (1+ \pi G  B^2 \rho^2 )^2 (-dt^2 +dz^2 +d \rho ^2 ) + {\rho ^2 d
\phi ^2 } (1 +  \pi G B^2 \rho ^2 )^{-2}
\eqno (47)
$$
The magnetic field is given by:
$$
F={{B \rho d \rho \wedge d \phi } \over {(1 + \pi G B^2 \rho ^2 )^2}}
\eqno (48)
$$

The Melvin solution possesses a degree of uniqueness.  For example Hiscock
has
shown the

\noindent {\bf Theorem:}  The only axisymmetric, static solution of the
Einstein-Maxwell field equations without an horizon which is
is asymptotically Melvin is in fact the Melvin Solution.

In fact Hiscock also allows for a neutral or electrically charged black hole as
well.

In fact one  can show that
the only translationally invariant static
solution  of the
Einstein-Maxwell field equations without horizon which is
asymptotically  Melvin is in fact the Melvin solution.

\noindent {\sl Proof}: assume the metric is static and has reflection
invariance with respect
to the $z-$direction. These two assumptions may easily be justified. The metric
takes the form
$$
ds^2=-V^2 dt^2 +Y^2 dz^2 +g_{AB}dx^A dx^B
$$
with $A=1,2$. The field equations are:
$$
\nabla ^A (VY \nabla _A  \ln (V/Y) )=VY8 \pi G( T_{ \hat z  \hat z } +
T_ { \hat 0 \hat 0 }   )
$$
$$
\nabla ^A (VY \nabla _A \ln (VY)) = VY 8 \pi G T ^A  _A
$$
$$
V ^{-1} \nabla _A \nabla _B V + Y^{-1} \nabla _A \nabla _B Y = K g_{AB} -8 \pi
(T_{AB} - {1 \over 2 } g_{AB}(T ^A _ A + T _{\hat z \hat z } + T _{ \hat 0
\hat 0  } ) )
$$
where $ K $ is the Gauss curvature of the 2-metric $g_{AB}$.
The electromagnetic field is assumed to be of the form:
$$
F= {1 \over 2 } F _{AB} dx ^A dx ^B .
$$
It follows that $T_ { \hat 0 \hat 0 } + T _{ \hat z \hat z } =0$ and hence:
$$
\nabla ^A ( VY \nabla _A (V/Y))=0.
$$
Now $V/Y$ tends to one at infinity (asymptotic boost invariance) and so we may
invoke tha Maximum Principle to show that $V=Y$ everywhere. Thus the metric
  must be boost invariant.It now follows that
$$
\nabla_A \nabla _B V = f g_{AB}
$$
for some scalar $f$. Thus
$$
K^A = \epsilon ^{AB} \nabla _B V
$$
is a Killing vector field of the 2-metric $g_{AB}$ and since
$K^A \partial _A V =0$
it is also a Killing vector field of the entire 4-metric.
It is not difficult to see
that this Killing vector field  corresponds to rotational symmetry of the
solution.

Having established the credentials of the Melvin solution as uniquely suitable
model of a static magnetic field in general relativity we turn to looking for
instanton solutions representing the creation of a black hole monopole
anti-monopole pair. If there were no external magnetic field the obvious
candidate instantons would be the magnetically charged C-metric for which
$$
G(x) = 1-x^2 -2GMA x^3 - G(g^2/4 \pi ) A^2 x^4 .
$$
However this has nodal singularities. In fact since  the metric is boost
invariant it has zero ADM mass and thus it cannot be regular by the positive
mass theorem generalised to include apparent horizons. However it was pointed
out by Ernst [18] that the nodal singularity may be eliminated if one appends a
suitable magnetic field. The resulting metric is of the same form as (38)
but the first three terms are mutiplied by and the last term divided
by the
factor:
$$ (1+GBgx/2)^4 .
$$
If $M=0=g=A$ we get the Melvin solution but the limit must be taken carefully.
The nodal singularity may be eliminated if $B$ is chosen so that
$$
G^{\prime}(x_3) /(1+GBgx_3/2)^4 \ \ + G^{\prime} (x_4)/(1+Ggx_4/2)^4 =0.
$$
This equation may be regarded as an equation for $B$ the magnetic field
necessary to provide the force to accelerate the magnetically charged black
hole. It is difficult to find an explicit solution in terms of $g$, $m$ and $A$
except when $GMA$ is small in which case one finds the physically sensible
result:
$$
gB \approx MA .
$$
In order to obtain an instanton which is regular on the Riemannian section
obtained by allowing the time coordinate $t$ to be pure imaginary it is
necessary that the $\tau = i t $ is periodic with period given by the surface
gravity. This leads to the condition that
$$
G^{\prime}(x_2) + G^{\prime}(x_3)=0.
$$
It appears that the the only way to satisfy this condition is to set:
$$
m=|g|/{\surd (4 \pi G})
$$
Note that this equation  implies that the horizons have a non-vanishing  common
surface gravity  and hence temperature as in the Mellor-Moss case.
It is not difficult to see
that the topology of the Riemann section is $S^2 \times S^2 $ with a point
(corresponding   to $x+y=0$ )
removed. In fact topologically one can obtain this manifold from ${\Bbb R}^4$, which
is the topology of the Melvin solution, by surgery along an
$ S^1$. That is by cutting out a neighbourhood of a circle which has topology
$D^3 \times S^1$ with boundary $ S^2 \times S^1$ and replacing by $S^2 \times
D^2$ which has the same boundary. This surgery is also what is needed to
convert
${\Bbb R}^3 \times S^1$ to ${\Bbb R}^2 \times S^2$ i.e. to convert a manifold
with the
topology of "Hot Flat Space" to that with the topology of the Riemannian
section of the Schwarzschild solution.  This apparent connection between surgery
along links and virtual black holes is an intriguing one and deserves to be
investigated in more detail.
The existence of the Melvi-Ernst
instanton would seem to be rather important. It seems to imply for
example that it would be
{\bf
inconsistent} not to consider the effects of black hole monopoles since given
strong enough magnetic fields they will be spontaneously created. Once they are
created they should evolve by thermal evaporation to the extreme zero
temperature soliton state. Another reason why I believe that this process is so
important is that it seems to show that while one may have one's doubts about
the effects of wormholes because of the absence of suitable solutions of the
classical equations of motion with positive definite signature, the solutions
described here do indicate that some sort of topological fluctuations in the
structure of spacetime {\bf must} be taken into account in a satisfactory
theory of gravity coupled to Maxwell or Yang-Mills theory.

I will now  sketch
how real tunneling geometries, which effectively exhaust the class of metrics
which allow a Wick rotation, are especially well adapted to implementing the
idea of Reflection Positivity used in flat space Euclidean Quantum Field
theory. These ideas are not new - they resemble some ideas of Uhlmann [19] and I
reviewed them briefly in my talk at the  Jena GRG conference [20]. however since
that time virtually nothing has been done (except [21]) on this. The time now
seems ripe for developing them and I understand from Bernard Kay that he and
Bob Wald also have some ideas in this direction.

The  main point made by Uhlmann is that the geometric data needed for
reflection positivity is a Riemannian manifold together with an isometric
involution $\theta$ having exactly the properties that I listed earlier, i.e.
such that the equations (18)-(21) hold. For {\bf
Euclidean}, i.e. flatspace,
Quantum Field theory of course the manifold is 4-dimensional Euclidean space
and $\theta$ is reflection in a hyperplane of constant imaginary time. Given
this data one may construct, without even passing to the associated Lorentzian
spacetime, the Hilbert Space of Quantum Mechanics in purely Riemannian
geometric language. The Riemannian manifold $2M_R$ may admit other isometries
in addition to $\theta$. If so the construction automatically builds in a
degree of equivariance with respect to those isometries. The "standard" case
is when $2M_R = S^4$ with its round metric. The isometry group is $O(5)$ and
the map $\theta$ commutes with an $O(4)$ subgroup which stabilizes $\Sigma$ as
a set. As mentioned above, if we assume that the isometry group acts
transitively on $2M_R$ the possibilities are quite limited by Ishihara's
results.

I shall confine attention to the case of a free massive scalar field with mass $m
>0$. We first construct the one particle Hilbert Space $ { \cal H}_1$. The
whole Hilbert space ${ \cal H}$ is built up by taking the direct sum of the
symmetric powers of ${ \cal H}_1$ under the tensor product. The k'th symmetric
power is the k-particle Hilbert Space. Thus:
$$
{ \cal H} ={\Bbb C} \oplus { \cal H}_1 \oplus { \cal H}_1 \otimes _S { \cal H}_1 \oplus
...
$$

I shall not dwell on function-analytic details so I will not specify very precisely
the function spaces and their completions. What I am interested in are the
basic geometric and physical ideas behind the construction. One begins by
identifying ${ \cal H}_1$ as a vector space with ${ \cal L}^2 (M^+_R)$, i.e.
with complex valued functions with support solely in $M^+_R$. One may think of
${ \cal H}_1$ as being made up of "positive frequency functions ". Recall  that
in flat Minkowski spacetime positive frequency functions may be characterized
as being holomorphic in the lower half complex t-plane. We have the obvious
orthogonal direct sum:
$$
{ \cal L}^2(2M_R) = { \cal L}^2 (M^+_R) \oplus { \cal L}^2 (M^-_R)
$$
where the $ { \cal L}^2$ norm is with respect to the Riemannian volume element
$\surd g d^4x$. The involution  $\theta$ acts on functions by pullback, i.e. if
$f^+(x) \in { \cal L}^2 (M^+_R)$ then $\theta ^* f^+(x)= f^+(\theta ^{-1} x) =
f^+(\theta x) \in { \cal L}^2 (M^-_R)$. Note that $\theta ^*$ is a selfadjoint
operator on ${\cal L}^2 (2M_R)$ which commutes with complex conjugation.
For notational convenience I will drop
the $*$ on $\theta ^*$ from now on . Some other useful notation is to define
the projections $\Pi _{\pm}$ onto ${\cal L}^2 (M_R^{\pm})$ and the projections
$P_{\pm}= {1 \over 2} (1 \pm \theta) $ onto even and odd functions with
respect to $\theta$. Thus $\Pi_{\pm} P_{-}$ projects onto functions on
$M^{\pm}_R$ satisfying
Dirichlet boundary conditions on $\Sigma$, that is to say they vanish on
$\Sigma$, while $\Pi_{\pm} P_+$ projects onto functions on $M^{\pm}_R$
satisfying Neumann boundary conditions on $\Sigma$. Although ${\cal L}^2
(M^+_R)$ comes equipped with its defining Hilbert metric this does not give the
correct norm for the one-particle Hilbert Space ${\cal H}_1$. To construct
this, which we write as $ \Vert f^+(x) \Vert ^2$, we need to introduce an
appropriate Green's function or two-point function, $G(x,y)$, on $2M_R \times
2M_R$. For a free scalar  field with mass $m$ we take the inverse of the
Klein-Gordon operator $ - \nabla ^2_{ g_R} + m^2 $  which is a positive self-adjoint on
${\cal L}^2(2M_R)$, where $\nabla ^2_{g_R}$ is the Laplacian with respect to
the Riemannian metric $g_R$ and has a unique inverse $G = (-\nabla ^2 _{g_R}
+m^2 )^{-1}$. Clearly $G$ commutes with  $\theta$. Two other Greens functions
are of interest. They are defined on ${\cal L}^2 (M^+_R)$ and satisfy
Dirichlet, $G_D$, or Neumann, $G_N$, boundary conditions. Thus:
$$
G_D = (1- \theta) G,
\eqno(49)
$$
i.e.
$$
G_D = G(x,y) -G(x,\theta y),
\eqno(50)
$$
and
$$
G_N =(1+\theta) G
\eqno (51)
$$
$$
  =G(x,y) +G(x,\theta y).
\eqno (52)
$$

We are now in a position to define $\Vert f^+(x) \Vert ^2$ as
$$
\Vert f^+(x) \Vert ^2 = \int _ {2M_R \times 2M_R}
{\overline {f^+(\theta x)}} G(x,y) f^+(y)
\eqno (53)
$$
$$
= \int _{M^-_R \times M^-_R}   {\overline {\theta f^+(x)}} G(x,y) f^+(y)
\eqno(54)
$$
$$
\int _{M^-_R} {\overline {f^-(x)} } \phi^+(x),
\eqno(55)
$$
where, $f^- =\theta f^+ \in {\cal L}^2 (M^+_R)$ and $\phi^+ \in {\cal L}^2
(2M_R)$ is the potential due to the source $f^+ \in {\cal L}^2 (M^+_R)$.

To justify the notation $\Vert f^+(x) \Vert ^2$ we must at least establish that
the right hand side of (5.6) is indeed positive. If we had chosen an arbitrary
two point function $G(x,y)$, even if it were pointwise positive such as the
Gaussian function in Euclidean 4-space, this would not
have been true so there is something non-trivial to be shown. The result
depends on some  special properties of the Klein Gordon operator. There are at
least two ways to proceed. One is to follow de Angelis et al. [21] and show by
means of Green's identity and some manipulations that :
$$
\int _{M^+_R \times M^+_R} {\overline {f^-}} \phi ^+
= 2 \int _{M^-_R} \vert \nabla \phi ^+ \vert
^2 + m^2 \vert \phi ^+ \vert ^2.
\eqno (56)
$$

Another way, following Glimm and Jaffe [22] and used by Uhlmann [19] is to make
use of the Dirichlet Principle. One re-writes (5.6) using (5.3) as
$$
\Vert f^+(x) \Vert ^2 = \int _{M^+_R} {\overline {f^+(x)}} \bigl( G(x,y) -
G_D (x,y)
\bigr) f^+(y).
\eqno(57)
$$
We may interpret $\Vert f^+(x) \Vert ^2$ in terms of a simple 4-dimensional
electrostatic model as the mutual potential energy of a
charge distribution located entirely in $M^+_R$  and given by $f^+(x) \in {\cal
L}^2 (M^+_R) $  with an image charge distribution  obtained by reflecting
$f^+(x)$ in the "conducting" hypersurface $\Sigma$ and taking the complex
conjugate. According to (5.10) this is the difference between two terms of the
form:
$$
\int _{M^+_R} \phi {\overline  {f^+(x)}}
\eqno(58)
$$
where $\phi$
satisfies
$$
\bigl( - \nabla ^2 _{g_R} + m^2 \bigl) \phi = f^+(x).
\eqno(59)
$$
The first term in (57) corresponds to  demanding as a boundary condition for
(59) that the extension of $\phi$
to $2M_R $ is in ${\cal L}^2 (2M_R)$ while the second term corresponds to
imposing the Dirichlet boundary condition on $\phi$. Dirichlet's Principle
states that among all solutions $\phi$ of (59), that satisfying Dirichlet
conditions has the least value for the integral (58). To prove this fact we
follow
Glimm and Jaffe and compare the positive and commuting operators $G^{-1}$
and $ G_D
^{-1}$. One may regard $G_D ^{-1}$ as the restriction of $ G_{-1}$ to functions
in ${\cal L}^2 (M^+_R)$ which in addition satisfy the condition that they
vanish on $\Sigma$. It follows that as operators on ${\cal L}^2 (M^+_R)$
$$
G_D \leq G
$$
and hence that $\Vert f ^+ (x) \Vert ^2 $ is indeed positive.
This second procedure is rather less direct than the one given previously. It
does however have the advantage that it generalizes to other situations. It is
can be used to obtain a proof in the case of flat Euclidean 4-space that the
generator of imaginary time translations is a positive operator which serves as
the physical Hamiltonian.

In the present case the Riemannian manifold $2M_R$ cannot be expected to admit
a translation Killing vector but it may well have some continous isometries
belonging to some group $K_R= {\rm Isom}_{0} (2M_R, g_R) $ the identity
component
of the isometry group.
The group $K$  will act on ${\cal L}^2 (2M_R)$ by pullback and via this the
analytic continuation $K_L$ should act on
the physical Hilbert space ${\cal H}$. In particular the physical
vacuum should be
invariant under $K_L$. A rather general discussion of this topic has been given
by [23] in the case that $2M_R$ is a symmetric space. This would include the most
important case which is $S^4$ or DeSitter spacetime. There seems little doubt
that the resulting vacuum state is the well known DeSitter invariant one,
although this has not, to my knowledge, ever been checked in full detail.

The
origin of DeSitter invariance described above is very similar to that given by
D'Eath  and Halliwell [24] in the context of Hawking and Hartle's  "No boundary
Proposal". In the more general context of tunneling transitions it suggests a a
natural candidate for the created quantum state after tunneling.
\vskip 1.5cm

\vfill
\vskip 2.0cm
\leftline {\bf References}
\item {[1]} R.P. Geroch, J. Math. Phys. {\bf 8} 782-786 (1968)
\item {[2]} G.W. Gibbons, Nucl. Phys {\bf B271} 479 (1986); N. Sanchez B.F. and
Whiting, Nucl. Phys. {\bf B283} 605 (1987)
\item {[3]} G.W. Gibbons and J.B. Hartle, Phys. Rev. {\bf D42} 2458 (1990)
\item {[4]} J.B. Hartle and S.W. Hawking, Phys. Rev. {\bf D28} 2960 (1983)
\item {[5]} J. Barrett, G.W. Gibbons, M.J. Perry and P.J. Ruback, in
preparation.
\item {[6]} M. Ferraris and M. Francaviglia, Gen. Relativ. Grav. {\bf 10} 283
(1979)
\item {[7]} D.J. Gross, M.J. Perry and L.G. Yaffe, Phy. Rev. {\bf D25} 330
(1982)
\item {[8]} G. Jungman and R.M. Wald, Phys. Rev. {\bf D40} 2615 (1989)
\item {[9]} R.L.  Bishop, Not. Amer. Math. Soc. {\bf 10} 364 (1963)
\item {[10]} S. Ishihara, J. Math. Soc. Japan {\bf 7} 345 (1955)
\item {[11]} M.J. Perry " An Instability of De Sitter Space" in
"The Very Early Universe"  edited by G.W. Gibbons, S.W. Hawking and S.T.C.
Siklos, Cambridge University Press (1983); P. Ginsparg and M.J. Perry, Nucl.
Phys. {\bf B222} 245 (1983)
\item {[12]} F. Mellor and I.G. Moss, Phys. Lett. {\bf B222} 361 (1989)
\item {[13]} G.W. Gibbons and C.M. Hull, Phys. Lett {\bf B 109} 190 (1982); G.W.
Gibbons, S.W. Hawking, G.T. Horowitz and M.J. Perry, Comm. Math. Phys. {\bf 88}
295 (1983)
\item {[14]} W Kinnersley and M Walker, Phys. Rev {\bf D2} 1359 (1970)
\item {[15]} G.W. Gibbons "Quantized Flux-Tubes in Einstein-Maxwell theory and
non-compact internal spaces" in Proceedings of the XIIth Karpac winter School
of Theoretical Physics: "Fields and Geometry" ed. A Jadczyk, World
Scientfic (Singapore)(1986)
\item {[16]} D. Garfinkle and A. Strominger, Phys. Lett. {\bf B256} 146 (1991)
\item {[17]} G.W. Gibbons " Self-gravitating magnetic monopoles, global
monopoles and black holes" to appear in Proceedings of the 1990 Lisbon Autumn
School on Physics edited by A.B. Henriques, Lecture Notes in Physics
(Springer-Verlag)
\item {[18]} F. J. Ernst, J. Math. Phys. {\bf 17} 515 (1976)
\item {[19]} A. Uhlmann,"On Quantization in Curved spcetime." in Proceedings of
the 1979 Serpukhov International Workshop on High Energy Physics ; Czech. J. Phys.  {\bf B31} 1249
(1981), {\bf B32 } 573 (1982); Abstracts of Contributed Papers to GR9 (1980)
\item {[20]} G.W. Gibbons "Quantization about Classical Background Metrics" in
Proceedings of the 9th G.R.G. Conference ed. E. Schmutzer, Deutscher Verlag der
Wissenschaften (1981)
\item {[21]} G. F. De Angelis, D. de Falco and G. Di Genova, Comm. Math. Phys.
{\bf 103} 297 (1985).
\item {[22]} J. Glimm and A. Jaffe, Lett. Math. Phys. {\bf 3} 377 (1979)
\item {[23]} J. Frohlich, K. Osterwalder and E. Seiler, Ann. Math. {\bf 118}
461 (1983)
\item {[24]} P.D. D'Eath and J.J. Halliwell, 

\bye